\documentclass{PoS}
\usepackage{amsmath}
\usepackage{wrapfig}
\usepackage{xcolor}

\title{Complex Langevin: Boundary terms and application to QCD}

\ShortTitle{CLE: Boundaries and QCD}

\author{\speaker{Manuel Scherzer}\\
        Institut f\"ur theoretische Physik, Universit\"at Heidelberg, Heidelberg, Germany\\
        E-mail: \email{scherzer@thphys.uni-heidelberg.de}}

\author{Erhard Seiler\\
        Max-Plack Institut f\"ur Physik (Werner-Heisenberg-Institut), Munich, Germany\\
        E-mail: \email{ehs@mpp.mpg.de}}

\author{D\'enes Sexty\\
Department of Physics, Bergische Universit\"at Wuppertal, Wuppertal, Germany\\
AS/JSC, Forschungszentrum J\"ulich, J\"ulich, Germany\\
E-mail: \email{sexty@uni-wuppertal.de}}

\author{Ion-Olimpiu Stamatescu\\
        Institut f\"ur theoretische Physik, Universit\"at Heidelberg, Heidelberg, Germany\\
        E-mail: \email{I.O.Stamatescu@thphys.uni-heidelberg.de}}

\abstract{We employ the Complex Langevin method for simulation of complex-valued actions. First, we show how to test for convergence of the method by explicitely computing boundary terms and demonstrate this in a model. Then we investigate the deconfinement phase transition of QCD with $N_f=2$ Wilson-fermions using the Complex Langevin Method and. We give preliminary results for the transition temperatures up to $\mu/T_c(\mu=0)\approx 5$ and compute the curvature coefficient $\kappa_2$.}

\FullConference{The 36th Annual International Symposium on Lattice Field Theory - LATTICE2018\\
		22-28 July, 2018\\
		Michigan State University, East Lansing, Michigan, USA.}

\begin{document}

\section{Brief introduction to Complex Langevin}
%\cms{1}\cehs{2}\cios{3}\cds{4}
The Complex Langevin equation models the Euclidean path integral measure $\rho=e^{-S}$ for some complex action $S\in\mathbb{C}$. This is done by complexifying the manifold ($\mathbb{R}\rightarrow\mathbb{C}$ or  $SU(N)\rightarrow SL(N,\mathbb{C})$)  and then setting up a stochastic process in the real and imaginary part of the complexified variable. In case of a simple model, this corresponds to 
\begin{align}
\frac{dx}{dt}&=-\text{Re}\left(\frac{dS(z)}{dz}\right)+\eta=K_x+\eta\nonumber\\
\frac{dy}{dt}&=-\text{Im}\left(\frac{dS(z)}{dz}\right)=K_y\, ,
\end{align}
where $z=x+iy$, $t$ is Langevin time and $\eta$ is a Gaussian random variable with mean zero and standard deviation $\sqrt{2}$. The process gives rise to an evolving probability measure $P(x,y;t)$. However, physics is described by the complex measure $\rho(z;t)\rightarrow e^{-S(z)}$ (for $t\rightarrow\infty$). This can be formalized \cite{Aarts:2009uq,Aarts:2011ax, Nagata:2016vkn} by demanding that
\begin{equation}
\left<\mathcal{O}(t)\right>_\rho=\int_\mathbb{R} \mathcal{O}(z)\rho(z;t) dz=\int_{\mathbb{R}^{2}}\mathcal{O}(x+iy)P(x,y;t)dxdy=\left<\mathcal{O}(t)\right>_P\, ,
\label{eq:EV}
\end{equation} 
for all times $t$, provided the initial conditions are chosen such that \eqref{eq:EV} is true at $t=0$. In order to prove said equivalence, one defines an interpolating quantity
\begin{equation}
F_\mathcal{O}(t,\tau)=\int P(x,y;t-\tau)\mathcal{O}(x+iy;\tau)dxdy\, ,
\end{equation}
which can be shown to have properties $F_\mathcal{O}(t,t)=\left<\mathcal{O}(t)\right>_\rho$ and $F_\mathcal{O}(t,0)=\left<\mathcal{O}(t)\right>_P$. Hence, \eqref{eq:EV} is fulfilled if 
\begin{equation}
\frac{\partial}{\partial \tau}F_\mathcal{O}(t,\tau)=0\, ,
\label{eq:dF}
\end{equation}
expressing the vanishing of certain boundary terms. Another problem of Complex Langevin simulations not discussed here is the problem of non-holomorphic actions, arising for instance by zeroes in the fermion determinant. The latter is discussed in \cite{Mollgaard:2013qra,Aarts:2017vrv,Bloch:2017sex}; a possibility to tackle the appearance of poles in QCD has been proposed in \cite{Nagata:2018mkb}. 

\section{Boundary terms for Complex Langevin}
So far, the argument has been rather mathematical. Here we derive an explicit quantity which can be computed from a Monte Carlo simulation to determine a posteriori whether the simulation results correspond to the correct expectation values $\left<\mathcal{O}\right>_\rho$ (a more detailed derivation can be found in  \cite{Scherzer:2018hid}). The starting point for this derivation is \eqref{eq:dF}, which can ultimately be simplified to \cite{Aarts:2009uq}
\begin{align}
\frac{\partial F_\mathcal{O}}{\partial \tau}=&\int\left[\left(\partial_y K_y\right)P(x,y;t-\tau)\right] \mathcal{O}(x+iy;\tau)dxdy\nonumber\\
+&\int P(x,y;t-\tau)K_y \partial_y\mathcal{O}(x+iy) dx dy \, .
\end{align} 
The integrand is a total derivative in $y$, hence the integration yields an explicit expression for the boundary term, if we introduce a finite cutoff $Y$ in the noncompact direction. This cutoff has to be taken to infinity numerically in the end.
The boundary term then reads
\begin{align}
B_\mathcal{O}(Y;t,\tau)=&\int [K_y(x,Y)P(x,Y;t-\tau)\mathcal{O}(x+iY;\tau)\nonumber\\
-&K_y(x,-Y)P(x,-Y;t-\tau)\mathcal{O}(x-iY;\tau)] dx\, .
\label{Bterm}
\end{align}
The evaluation of the boundary term is practically impossible for theories other than toy models as it involves the calculation of $\mathcal{O}(x,y,tau)$ for $\tau>0$.
Generally in a compact space the observables will be of the form
\begin{equation}
\mathcal{O}=\sum_k c_k e^{ikx}\, ,
\end{equation}
where now the Fourier coefficients carry the time evolution. To gain more insight on how to simplify this we investigate a simple model
\begin{equation}
S(x)=i\beta \text{cos}(x) \, ,
\label{eq:action}
\end{equation}
which has been shown \cite{Salcedo:2016kyy} to lead to the following the stationary probability distribution under Complex Langevin evolution
\begin{equation}
P(y)=\frac{1}{4\pi \text{cosh}^2(y)}\, .
\end{equation}
It turns out that this does not give the correct results following $\rho(z)=e^{-S(z)}$.
Since this model is simple, one can explicitely solve the Fokker-Planck equation of the system to see what goes wrong.
\begin{figure}
\centering
\includegraphics[width=0.45\textwidth]{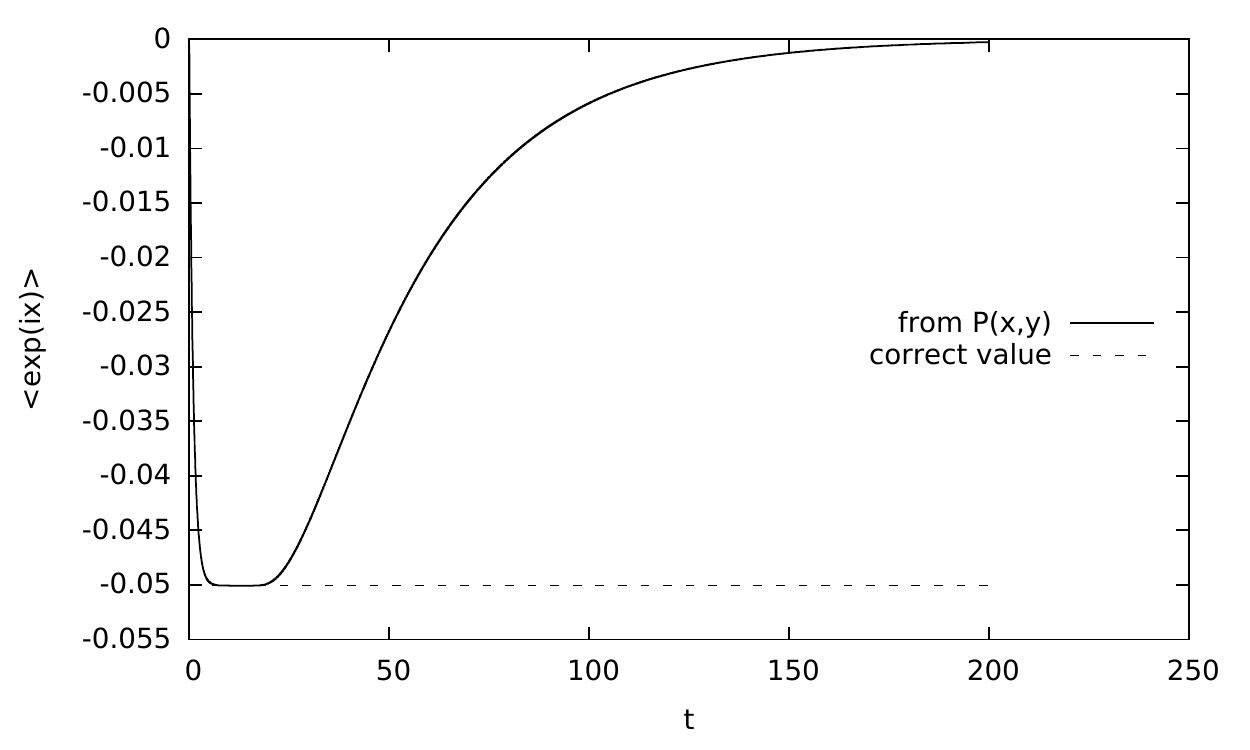}
\includegraphics[width=0.45\textwidth]{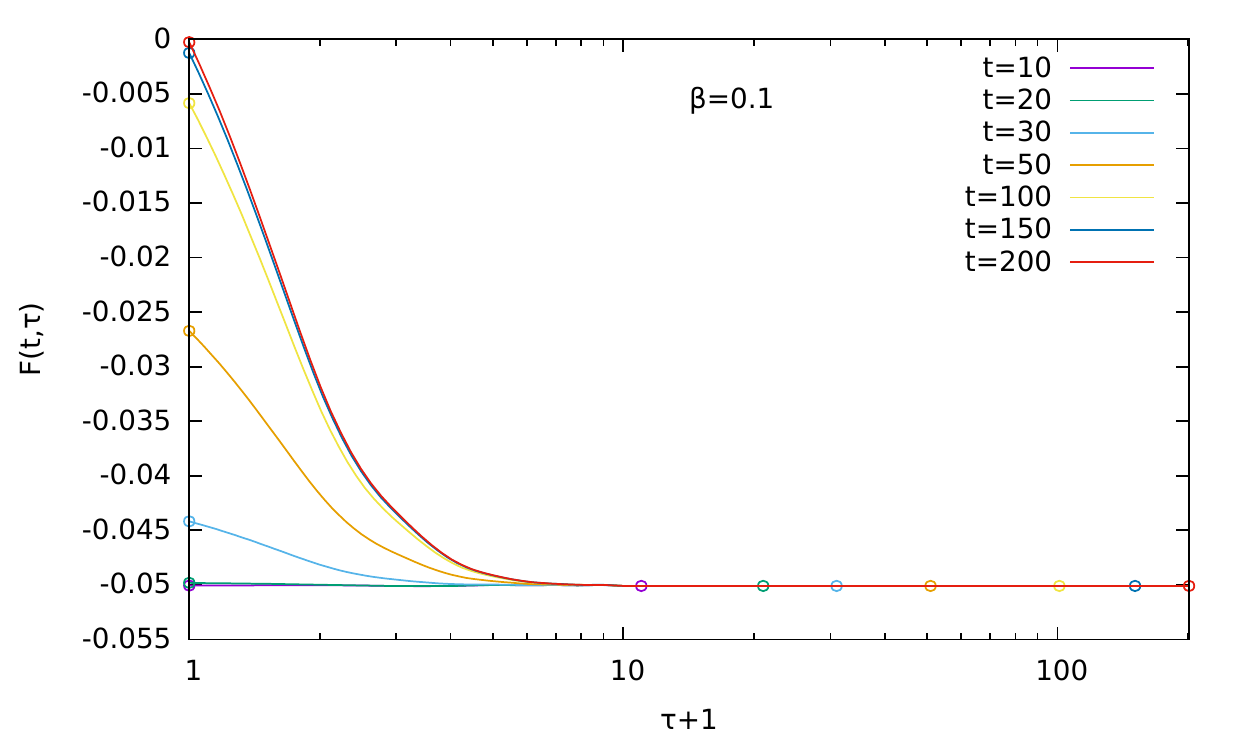}
\caption{Left: Time evolution of $\mathcal{O}_1=e^{ix}$; Right: $F_1(t,\tau)$ vs. $\tau$ for different $t$. Both figures are for $\beta=0.1$.}
\label{fig:FPevol}
\end{figure}
From this solution one can draw some conclusions, which are depicted in fig. \ref{fig:FPevol}. (1) initially the evolution goes to the correct value but then starts to deviate around $t~20$; (2) $\partial_{\tau} F_\mathcal{O}(t,\tau)$ appears to be maximal at $\tau=0$ for times $t$ after the plateau, which suggests that the effect of possible boundary terms is the strongest at that point. Taking (2) to heart and setting $\tau=0$ in \eqref{Bterm} makes the numerical evaluation much simpler, since at $\tau=0$ the observable is just given by the initial condition for the evolution ofthe observable. For the model \eqref{eq:action} using $\mathcal{O}=\text{exp}(ikx)$ and taking $t\rightarrow \infty$ the boundary term becomes
\begin{equation}
B_1(Y;\infty,0)=-\frac{i\beta}{2}\text{tanh}(Y)\, .
\end{equation}

\begin{wrapfigure}{l}{0.5\textwidth}
\centering
\includegraphics[width=0.45\textwidth]{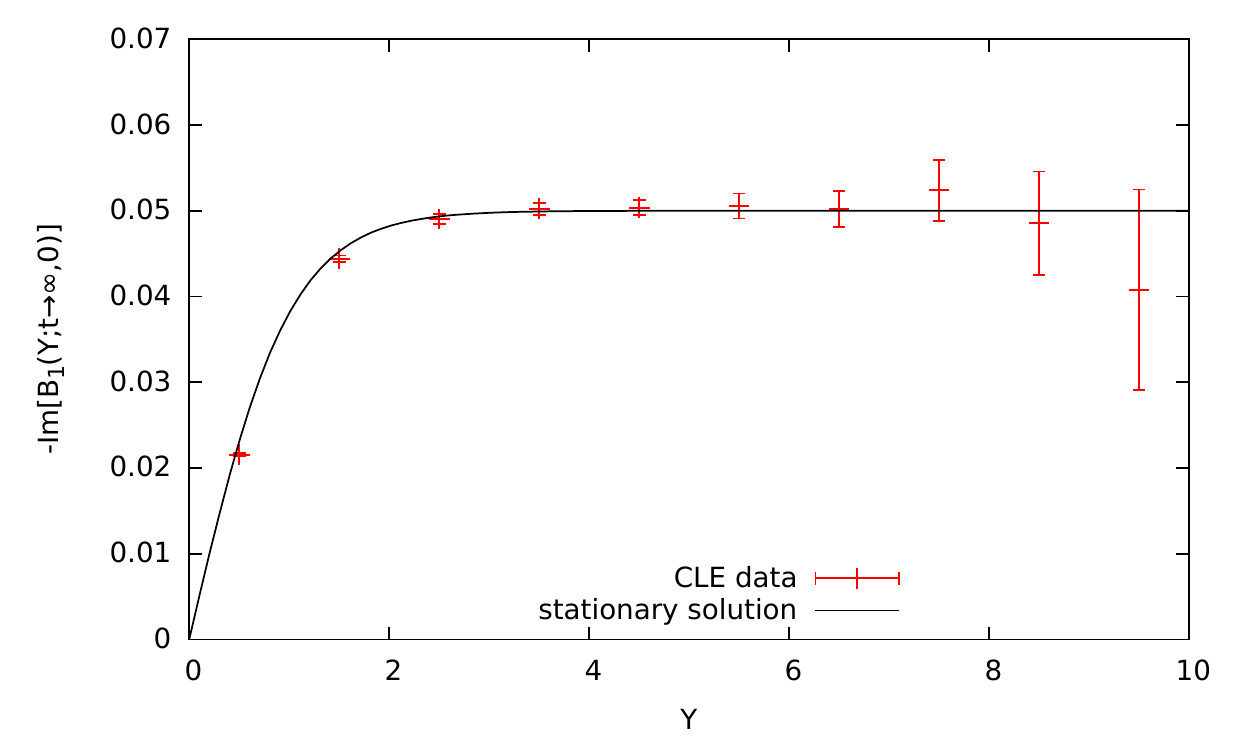}
\caption{Boundary term as a function of $Y$ from a Complex Langevin simulation.}
\label{fig:bound}
\end{wrapfigure}

$B_1(Y;\infty,0)$ can also be evaluated within a Complex Langevin simulation. This is done by simply doing a binning in $Y$ during the simulation and then computing \eqref{Bterm}.
The result is shown in fig.~\ref{fig:bound}.\\
We are currently working on extending this criterion to higher dimensional spaces (especially lattice simulations), where it is also possible to compute the boundary term explicitely. Hence, this is a definite criterion which can be used to determine a posteriori whether the result of a Complex Langevin simulation can be trusted.
Generally those boundary terms appear due to the simulation travelling to far in the noncompact direction.
This can be often be avoided by gauge fixing or gauge cooling \cite{Seiler:2012wz} to yield correct results. 
Adding a regularisation term damping the drift in the imaginary direction also helps \cite{Loheac:2017yar} but needs extrapolation to the non-regularized model \cite{Attanasio:2018rtq}.

\section{The deconfinement transition from Complex Langevin}
\subsection{Complex Langevin for QCD}
\begin{wrapfigure}{l}{0.5\textwidth}
\centering
\includegraphics[width=0.45\textwidth]{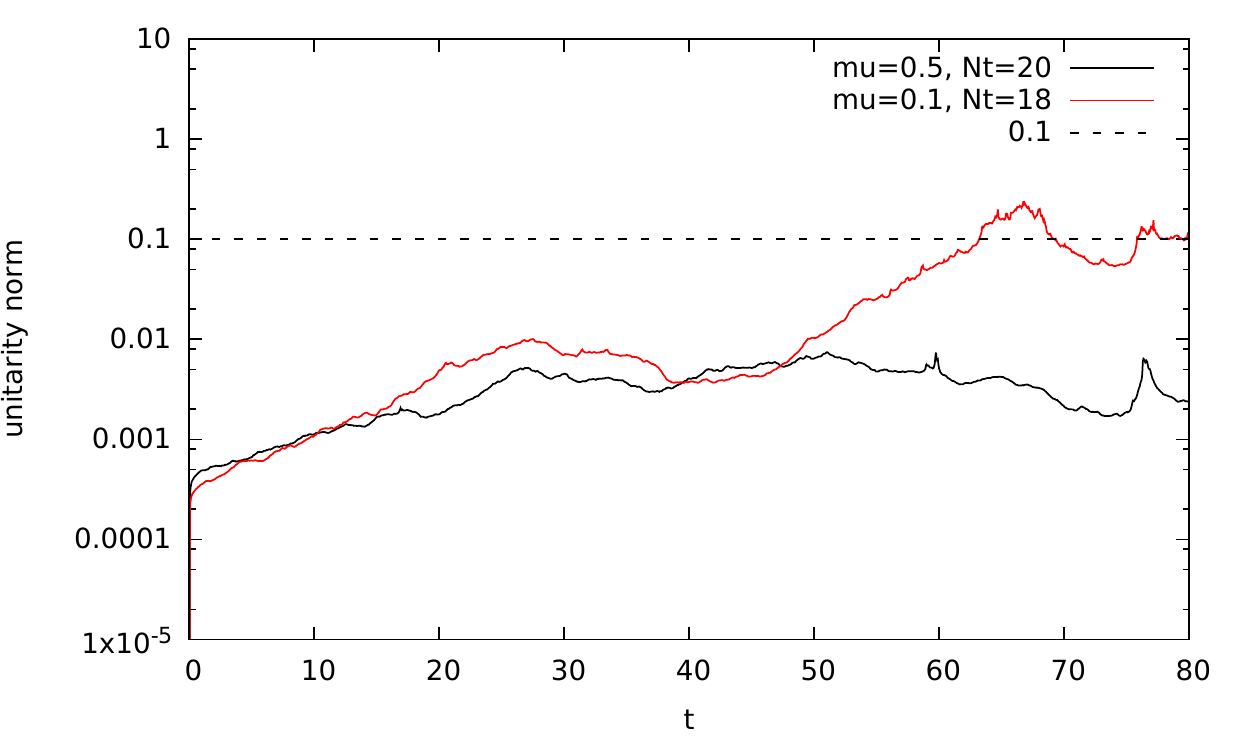}
\caption{Two examples for typical evolutions of the unitarity norm. If it rises above $0.1$, we choose to discard the following values. At which point in $t$ is rises above this value is mostly random.}
\label{fig:unorm}
\end{wrapfigure}

So far we have looked at validity and convergence of the Complex Langevin method. Ultimately we are interested in application to more interesting theories, the most interesting of which arguably is QCD. Here we adopt the Wilson fermion discretization without improvement and simulate lattice QCD with $N_f=2$ fermion flavors, for a recent investigation see also \cite{Aarts:2014bwa}.
 We simulate at coupling $\beta=5.9$ and hopping $\kappa=0.15$, where the mass is large enough (we have $m_\pi\sim2.1\text{GeV}$) for zeroes of the determinant not to show up. The results presented here are for a low volume $N_s=8$, but we are currently simulating larger volumes. 
The Complex Langevin updating prescription for QCD reads \cite{Sexty:2013ica}
\begin{equation}
U_{x,\nu}(t+\epsilon)=\text{exp}\left[i\sum_a \lambda_a\left(\epsilon K_{a,x,\nu}+\sqrt{\epsilon}\eta_{a,x,\nu}\right)\right]U_{x,\nu}(t)\, ,
\end{equation}
with the Gell-Mann matrices $\lambda_a$, Gaussian noise with zero mean and $\sqrt{2}$ standard-deviation and the Langevin drift
$K_{a,x,\nu}=-D_{a,x,\nu}S_\text{YM}+N_F \text{Tr}\left[M^{-1}(\mu,U)D_{a,x,\nu}M(\mu,U)\right]$
with the Wilson fermion matrix M and the derivative defined as $D_{a,x,\nu}f(U)=\partial_\alpha f(e^{-i\alpha\lambda_a}U_{x,\nu})|_{\alpha=0}$. As discussed in the last section, Complex Langevin simulation typically are plagued by the appearance of boundary terms. QCD is no exception to this rule. One can measure how far the simulation travels in the noncompact direction by means of the unitarity norm \cite{Aarts:2008rr}
\begin{equation}
N_U=\left<\text{Tr}\left(U^\dagger U - I\right)\right>_{x,\nu}\, ,
\label{eq:unorm}
\end{equation}
where $I$ is the unit matrix. If the evolution stays in $SU(3)$ the unitarity norm is zero. It becomes larger the further the evolution departs from $SU(3)$. The unitarity norm can be controlled by employing gauge cooling \cite{Seiler:2012wz}, which simply consists of gauge transformations in the opposite direction of the gauge gradient of the unitarity norm \eqref{eq:unorm}.  A different approach to controlling the unitarity norm, called ``dynamical stabilisation'' \cite{Attanasio:2018rtq} is not discussed here. A typical evolution of the unitarity norm under gauge cooling can be seen in fig. \ref{fig:unorm}.  As one can see, even with gauge cooling the unitarity norm rises. Typically the evolution starts to deviate from the correct value, if $N_U>0.1$,
 which will be reached in most simulations. However, as long as all observables we are interested in thermalize and give sufficient statistics before the unitarity norm becomes too large, one still can extract physical results. This is similar to the plateaus that occured in the model in the previous section: As long as one only takes expectation values while the evolution is on the plateau, the results will be correct. In QCD this leads to many repeated shorter simulations being averaged over.\\~\\

\subsection{Determination of $T_c(\mu)$}
\begin{figure}
\centering
\includegraphics[width=0.45\textwidth]{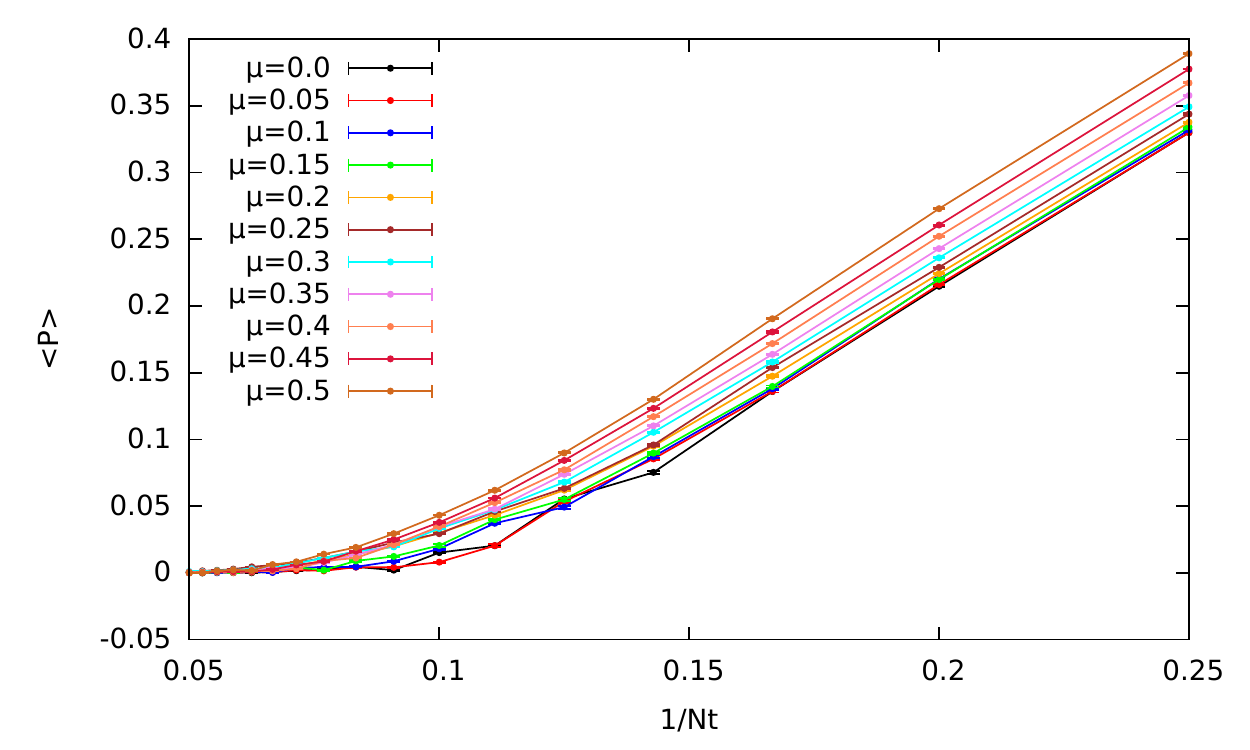}
\includegraphics[width=0.45\textwidth]{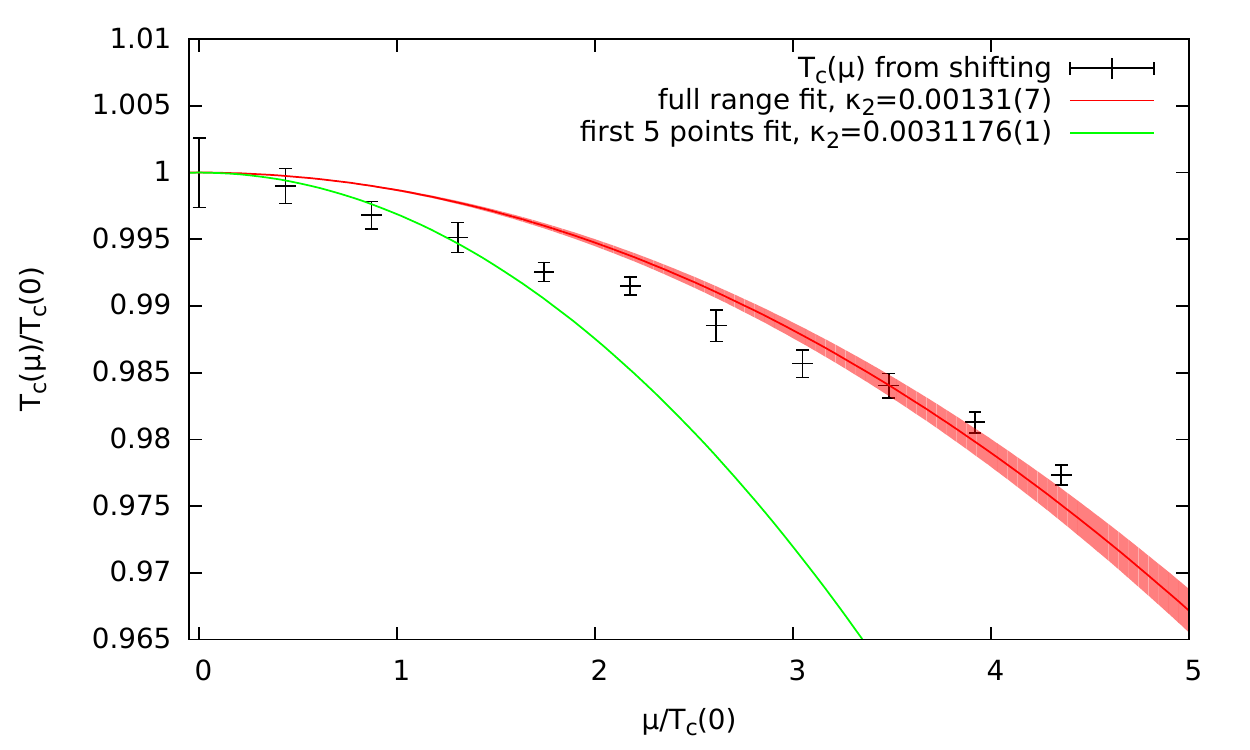}
\caption{Left: (unrenormalized) Polyakov loop vs. of $1/Nt$; Right: $T_c(\mu)$ transition line and quadratic fit.}
\label{fig:ploop}
\end{figure}

We determine the transition line of QCD by keeping the parameters of the action $\beta$, $\kappa$, $\mu$ as well as the spatial volume fixed and to vary the temporal lattice extent in order to vary the temperature. This is done for several values of the chemical potential $\mu$, such that in the end we are able to determine the transition temperature as a function of chemical potential, $T_c(\mu)$. The observable we choose to characterize the crossover is the Polyakov loop
\begin{equation}
P_x=\text{Tr}\left[\prod_{i=0}^{N_t-1} U_{(x,t=i), \mu=4} \right]
\end{equation} which is the order parameter for the deconfinement transition in pure Yang-Mills theory and indicates a crossover in full QCD similar to the chiral condensate. Typically in Taylor expansion and analytic continuation approaches another observable, the chiral condensate is used, since it is less noisy than the Polyakov loop. Here we choose the Polyakov loop because the chiral condensate susceptibility shows a rather weak signal, which is possibly due to the rather large mass of our quarks.
The Polyakov loop is depicted in fig.~\ref{fig:ploop}  as a function of $1/N_t$ for different values of $\mu$.

Here, one can observe that the Polyakov loop $\left<P(T)\right>$ has the same form as and can be related to $\left<P(T=0)\right>$ by a simple shift in $T$.  Hence, we adapt an approach similar in spirit to the one presented in \cite{Endrodi:2011gv}. Given the Polyakov loop at some $\mu$,
$\left<P(T,\mu)\right>$, we choose $P(0)$ as a reference and determine a shift in temperature by minimizing
\begin{equation}
\left| \left<P(T,0)\right>- \left<P(T-\tau, \mu)\right>\right|
\end{equation}
with respect to $\tau$. I.e.~we compute how far in $T$ the Polyakov loop has to be shifted to match the curve at $\mu=0$. Provided the transition temperature at $\mu=0$ is known, one can compute the transition temperature as a function of $\mu$ via
\begin{equation}
T_c(\mu)=T_{c}(0)-\tau \, .
\end{equation}
The curvature coefficient $\kappa_2$ is then extracted by fitting 
\begin{equation}
\frac{T_c(\mu)}{T_c(0)}=1-\kappa_2 \left(\frac{\mu}{T_c(0)}\right)^2 \, 
\label{eq:fitfunc}
\end{equation}
to the data. This ansatz is usually used in Taylor expansion and analytic continuation approaches to determine the curvature of the transition line at $\mu=0$ \cite{Bellwied:2016cpq, Bazavov:2017tot, Bonati:2018nut}. Our approach differs from that definition in the sense that we do not only determine the curvature at $\mu=0$ but in a certain range of $\mu$.\\
In this work we use the bare Polyakov loop. This will not give the correct physics in the continuum and thermodynamic limit.
The Polyakov loop renormalizes multiplicatively as \cite{Borsanyi:2012uq, Bruckmann:2013oba}
\begin{equation}
P_\text{ren}(N_t,\mu)=\left(\frac{P^*}{P_\text{ref}(N_t^*,\mu=0)}\right)^{\frac{N_t}{N^*_t}} P_\text{bare}(N_t,\mu)\, ,
\end{equation}
where the idea is to set the $P_\text{ref}(N_t^*,\mu=0)$ to some reference value $P^*$. A good choice is to set the Polyakov Loop to $P^*=1$ at high temperature, which corresponds to free energy $F=0$. We postpone the analysis with proper renormalization to future work.\\
Alternatively a good way to extract the transition temperature is by means of the Binder cumulant \eqref{eq:Binder}, where multiplicative renormalization drops out.

We determine the transition temperature at $\mu=0$ by finding the inflection point of the Binder cumulant 
\begin{equation}
\frac{\left<P^4\right>}{\left<P^2\right>^2}\, ,
\label{eq:Binder}
\end{equation}
which interpolates between the value 3 in the confined phase and 1 in the deconfined phase.
Typically one would extract the transition temperature from the susceptibility, but since we only have datapoints at integer values of $N_t$, there are not enough points in the transition region, which makes it impossible to fit a peak to extract a good value for $T_c(0)$.
We define the crossover temperature as the inflection point of the Binder cumulant fitted by a logistic curve, giving a transition temperature of
\begin{equation}
T_c(0)=0.115(2)\, .
\end{equation}
Combining this with \eqref{eq:fitfunc} allows us to determine $T_c(\mu)$, which is depicted in fig.~\ref{fig:ploop}. The fit was done over two different region (R1) the full range and (R2) only the first 5 points ($\mu/T_c(0)<2$). The resulting curvatures are
\begin{equation}
\kappa_2^{\text{R1}}=0.0013139(652)\text{ and }\kappa_2^\text{R2}=0.0031176(1)\, .
\end{equation}
Note that those do not agree within errorbars, which suggests that the actual functional form of the transition is not quadratic.

{\em Acknowledgments:} M.~S., E.~S. and I.-O.~S. gratefully acknowledge 
kind support from DFG under Grant Sta 283/16-2. D.~S. gratefully 
acknowledges funding by the DFG grant Heisenberg Programme (SE 2466/1-2).
The authors acknowledge support by the High Performance and 
Cloud Computing Group at the Zentrum f\"ur Datenverarbeitung of the 
University of T\"ubingen, the state of Baden-W\"urttemberg through bwHPC
and the German Research Foundation (DFG) through grant no INST 37/935-1 FUGG.

\end{document}